# Finding meteorite impacts in Aboriginal oral tradition

Duane W. Hamacher

Nura Gili Indigenous Programs Unit, University of New South Wales, Sydney, NSW, 2052, Australia
Email: d.hamacher@unsw.edu.au

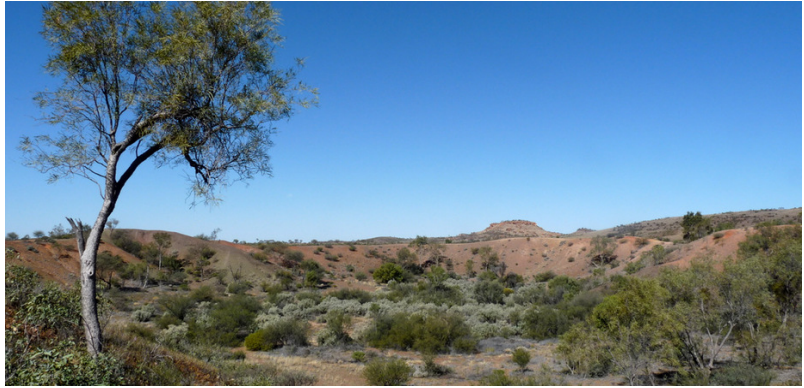

*Figure 1: Aboriginal stories dating back many thousands of years talk of a fire from the sky in an area now home to the Henbury meteorite craters, in the Northern Territory. Image: Flickr/Boobook, CC BY-NC-SA*

Imagine going about your normal day when a brilliant light races across the sky. It explodes, showering the ground with small stones and sending a shock wave across the land. The accompanying boom is deafening and leaves people running and screaming.

This was the description of an incident that occurred over the skies of **Chelyabinsk**, Russia on February 15, 2013, one of the best recorded meteoritic events in history. This airburst was photographed and videoed by many people so we have a good record of what occurred, which helped explain the nature of the event.

But how do we find out about much older events when modern recordings were not available?

A century before Chelyabinsk, a similar event occurred on July 30, 1908, over the remote Siberian forest near **Tunguska**.

That explosion was even more powerful, flattening 80 million trees over an area of 2,000 square kilometres and sending a shock wave around the Earth – twice. It was 19 years before scientists reached the Tunguska site to study the effects of the blast.

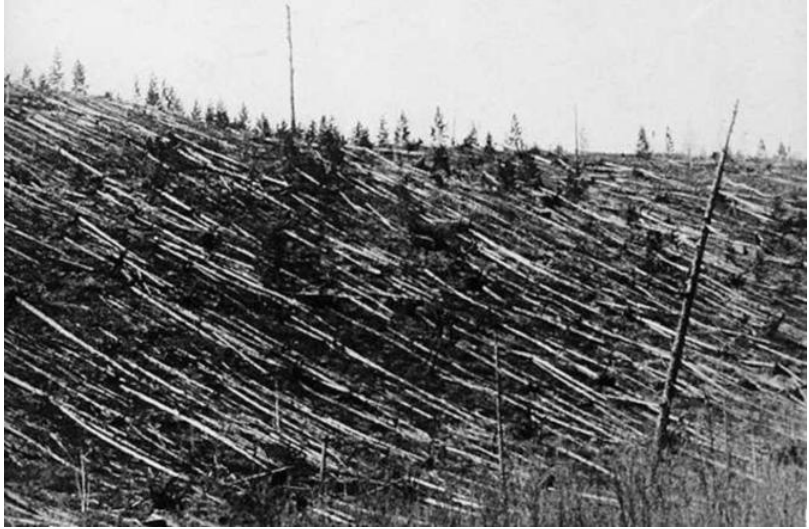

*Figure 2: Effects of the Tunguska blast 19 years after the event. Some of the trees flattened by the airburst can still be seen to this day. Image: Leonid Kulik*

The apparent lack of a meteorite fuelled speculation about how it formed, from sober suggestions of an exploding comet to more outlandish claims of mini-black holes and crashed alien spacecraft (**research confirms** it was an exploding meteorite).

**Meteoric events in Indigenous oral tradition**

In 1926, the ethnographer Innokenty Suslov interviewed the local Indigenous Evenk people, who still vividly remembered the Tunguska airburst.

At the time, a great feud persisted among Evenki clans. One clan called upon a **shaman** named Magankan to destroy their enemy. On the morning of July 30th, 1908, Magankan sent Agdy, the god of thunder, to demonstrate his power.

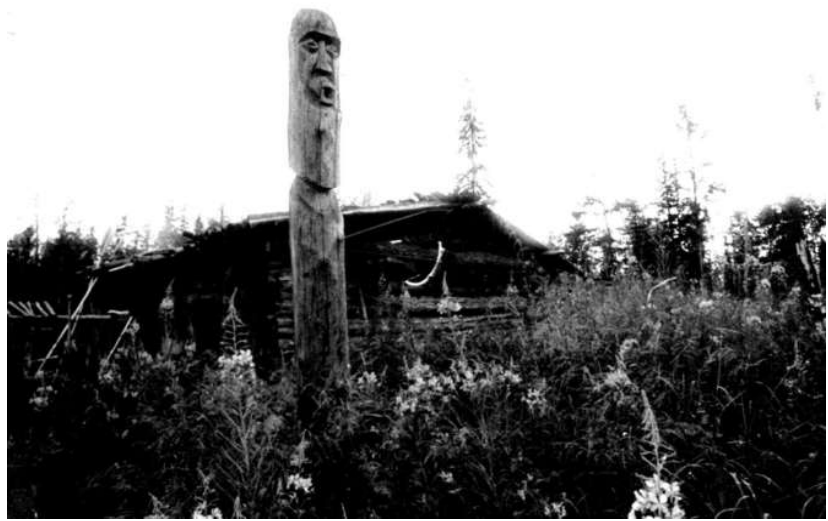

*Figure 3: A carving of the thunder god Agdy at Tunguska. Image: University of Bologna, Department of Physics.*

Many Indigenous cultures **attribute meteoritic events** to the power of sky beings. The Wardaman people of northern Australia tell of Utdjungon, a being who lives in the Coalsack nebula by the Southern Cross.

He will cast a fiery star to the Earth if laws and traditions are not followed. The falling star will cause the earth to shake and the trees to topple.

Like the Evenki, it seems the Wardaman have faced Utdjungon's wrath before.

The Luritja people of Central Australia also tell of an object that fell to Earth as punishment for breaking sacred law. And we can still see the scars of this event today.

**A surviving meteorite impact legend**

Around 4,700 years ago, a large nickel-iron meteoroid came blazing across the Central Australian sky. It broke apart before striking the ground 145km south of what is now Alice Springs.

The fragments carved out more than a dozen craters up to 180 meters across with the energy of a small nuclear explosion.

Today, we call this place the **Henbury Meteorites Conservation Reserve**.

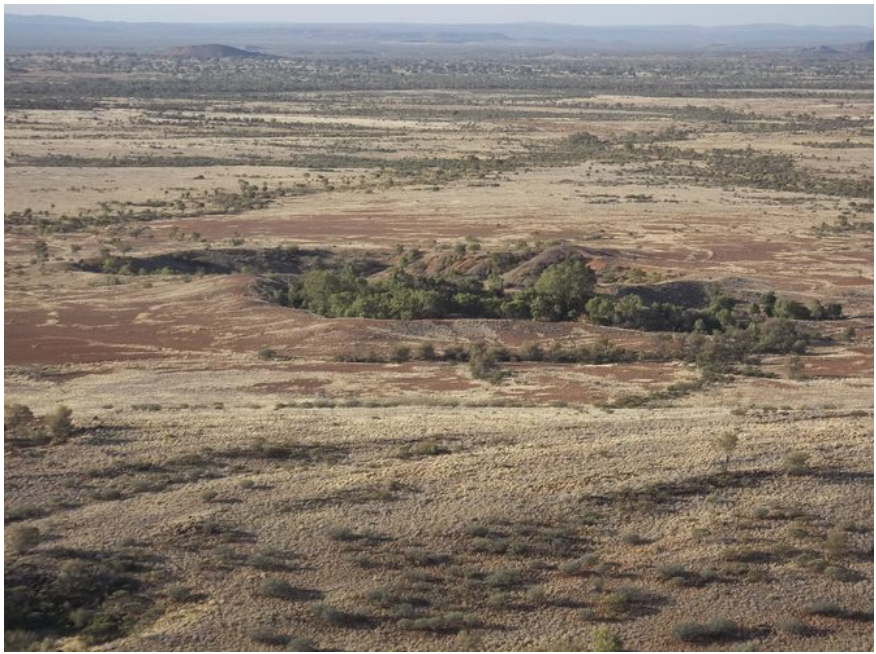

*Figure 4: A cluster of the largest craters at Henbury, as seen from the nearby Bacon Range. Image: Duane Hamacher*

Aboriginal people have inhabited the region for tens-of-thousands of years, and it's almost certain they **witnessed this dramatic event**. But did an oral record of this event survive to modern times?

When scientists first visited Henbury in 1931, they brought with them an Aboriginal guide. When they ventured near the site, the guide would go no further.

He said his people were forbidden from going near the craters, as that was where the fire-devil ran down from the sun and set the land ablaze, killing people and forming the giant holes.

They were also forbidden from collecting water that pooled in the craters, as they feared the fire-devil would fill them with a piece of iron.

The following year, a local resident asked Luritja elders about the craters. The elders provided the same answer and said the fire-devil "will burn and eat" anyone who breaks sacred law, as he had done long ago.

**The longevity and benefits of oral tradition**

The story of Henbury indicates a living memory of an event that occurred a few thousands of years ago. Might then we find accounts of events from tens of thousands of years ago?

Yes, it seems so.

Recent studies show that Aboriginal traditions accurately record **sea level changes** over the past 10,000 years.

**Other studies** suggest the volcanic eruptions that formed the Eacham, Euramo and Barrine crater lakes in northern Queensland more than 10,000 years ago are recorded in oral tradition.

In addition to demonstrating the longevity of Indigenous oral traditions, emerging research shows that these stories can lead to new scientific discoveries. **Aboriginal stories** about objects falling from the sky have led scientists to meteorite finds they would not have known about otherwise.

In New Zealand, **geologists** are also using Maori oral traditions to study earthquakes and tsunamis. New Zealand has a much more recent human history – compared to Australia – with the first Maori ancestors **thought to have arrived** around the 13th Century.

The arrival of the first Australians goes back **at least 50,000 years**. There is still much to learn, as Australia's ancient landscape has been exposed to meteorite strikes that we don't know about, some of which have probably occurred since humans arrived.

But given that Australia is home to the oldest continuing cultures on Earth, we are only just scratching the surface of the vast **scientific knowledge** contained in Indigenous oral traditions.

We anticipate that **our work** with Aboriginal elders to learn about **Indigenous astronomy** will lead to new knowledge and **cultural insights** about natural events and meteorite impacts in Australia.

**References and Further Reading**

---------------------------

*This article was originally published in The Conversation on 4 March 2015.*

*Dr Duane W. Hamacher is a Lecturer and ARC Discovery Early Career Research Fellow in the Nura Gili Indigenous Programs Unit at the University of New South Wales, in Sydney, Australia. His teaching and research focus on cultural astronomy and Indigenous Knowledge systems.*

*He receives funding from the Australian Research Council.*

---------------------------